\documentclass[12pt,a4paper]{article}
\usepackage{amsmath}
\usepackage{amsfonts}
\usepackage{amssymb}
\usepackage{makeidx}
\usepackage[applemac]{inputenc}

\begin{document}
\sloppy
\newtheorem{hypothesis}{Hypothesis}
\newtheorem{comment}{Comment}

\title{Paths and stochastic order in open systems}
\author{Umberto Lucia\\I.T.I.S. `A. Volta' \\Spalto Marengo 42, 15121 Alessandria, Italy}
\date{}
\maketitle

\begin{abstract}
The principle of maximum irreversible is proved to be a consequence of a stochastic order of the paths inside the phase space; indeed,  the system evolves on the greatest path in the stochastic order. The result obtained is that, at the stability, the entropy generation is maximum and, this maximum value is consequence of the stochastic order of the paths in the phase space, while, conversely, the stochastic order of the paths in the phase space is a consequence of the maximum of the entropy generation at the stability.
\end{abstract}

\section{Introduction}
It has been proved that entropy is the quantity which allows us to describe the progress of non-equilibrium dissipative process [\ref{1},\ref{2}]. Using the maximum entropy production principle, MEPP, it has been proved that a non-equilibrium system develops following the thermodynamic path which maximises its entropy generation under present constraints [\ref{Martyushev Seleznev  2006},\ref{Lucia 1995}]. Considering the statistical interpretation of entropy, entropy not only tends to increase, but it will increase to a maximum value. Consequently, MEPP may be viewed as the natural generalisation of the Clausius-Boltzmann-Gibbs formulation of the second law [\ref{Martyushev Seleznev  2006}]. Moreover, from 1995 to  2007 the principle of \textit{maximum entropy generation}, elsewhere called \textit{maximum entropy variation due to irreversibility} or \textit{maximum irreversible entropy} $S_{irr}$, has been proved for open systems [\ref{Lucia 1995}-\ref{Lucia 2008}].  The principle of maximum for the variation of the entropy due to irreversibility represents a general principle of investigation for the stability of the open systems; indeed, it states that: \textit{in a general thermodynamic transformation, the condition of the stability for the open system steady states consists of the maximum for the variation of the entropy due to the irreversibility}. This statement represents an important result in irreversible thermodynamics because it is a global theoretical principle for the analysis of the stability of open systems.  The irreversible entropy maximum principle represents the physical foundation of the design principle of Nature; indeed, in phenomena out of equilibrium, irreversibility manifests itself because the fluctuations of the physical quantities, which bring the system apparently out of stationarity, occur symmetrically about their average values and entropy generation represents the physical quantity which allows us to describe the stability and the range of the fluctuation around the stable states [\ref{Grazzini Lucia 2008}-\ref{Gallavotti 2006}]. It allows us to link the global approach to the statistical one when irreversible phenomena occur and to obtain the definition of the $PA$-measure [\ref{Lucia 2008}] for open systems: it represents the relation between the probability measurement in the probability space and real statistical facts. Using entropy generation it has been introduced the global thermodynamic effect of the action of the forces finding a global expression which links the microscopical values to the macroscopical thermodynamic quantities [\ref{Lucia 2008}].

A path information has been defined in connection with the different possible paths of chaotic system moving in its phase space between two cells. On the basis of the assumption that the paths are differentiated by their actions, Wang showed that the maximum path information leads to a path probability distribution as a function of action. An interesting result is that the most probable paths are just the paths of least action [\ref{Wang 2005}]. This suggests that the principle of least action, in a probabilistic situation, is equivalent to the principle of maximization of information or uncertainty associated with the probability distribution. Here we link this result with the principle of maximum entropy generation. The aim of this paper is to introduce the stochastic order in the analysis of the probability distribution attributed to the different paths of chaotic systems moving between two points in the phase space.

In this paper we prove that the principle of maximum irreversible entropy is related to the stochastic order of the paths inside the phase space. To do so, in Section 2 it will be introduced the system considered and the relation of the entropy generation to the theory of probability, in Section 3 path information is linked to irreversible entropy and in Section 4 paths are analyzed using the stochastic order.

\section{The open irreversible system}
In this section it will be defined the system considered [\ref{Lucia 1995}]. Let us consider an open continuum or discrete $N$ particles system. Every $i-$th element of this system is located by a position vector $\mathbf{x}_{i}\in \mathbb{R}^{3}$, it has a velocity $\Dot{\mathbf{x}}_{i}\in \mathbb{R}^{3}$, a mass $m_{i}\in \mathbb{R}$ and a momentum $\mathbf{p}=m_{i}\mathbf{\Dot{x}}_{i}$, with $i\in [1,N]$ and $\mathbf{p}\in \mathbb{R}^{3}$. The masses $m_{i}$ must satisfy the condition:
	\begin{equation}
		\sum_{i=1}^{N}m_{i}=m
	\end{equation} 
where $m$ is the total mass which must be a conserved quantity so that it follows:
	\begin{equation}
		\Dot{\rho}+\rho \nabla\cdot\Dot{\mathbf{x}}_{B}=0
	\end{equation} 
where $\rho =dm/dV$ is the total mass density, with $V$ total volume of the system and $\Dot{\mathbf{x}}_{B}\in \mathbb{R}^{3}$, defined as $\Dot{\mathbf{x}}_{B}=\sum_{i=1}^{N}\mathbf{p}_{i}/m$, velocity of the centre of mass. The  mass density must satisfy the following conservation law:
	\begin{equation}
		\Dot{\rho}_{i}+\rho_{i} \nabla\cdot\Dot{\mathbf{x}}_{i}=\rho \Xi
	\end{equation} 
where $\rho_{i}$ is the density of the $i-$th elementary volume $V_{i}$, with $\sum_{i=1}^{N}V_{i}=V$, and $\Xi$ is the source, generated by matter transfer, chemical reactions and thermodynamic transformations. A dynamical state of $N$ particles can be specified by the $3N$ canonical coordinates $\left\{\mathbf{q}_{i}\in \mathbb{R}^{3},i\in \left[1,N\right]\right\}$ and their conjugate momenta $\left\{\mathbf{p}_{i}\in \mathbb{R}^{3},i\in \left[1,N\right]\right\}$. The $6N-$dimensional space spanned by $\left\{\left(\mathbf{p}_{i},\mathbf{q}_{i}\right),i\in \left[1,N\right]\right\}$ is called the phase space $\Omega$. A point $\mathbf{\sigma}_{i} =\left(\mathbf{p}_{i},\mathbf{q}_{i}\right)_{i\in \left[1,N\right]}$ in the phase space $\Omega :=\left\{\mathbf{\sigma}_{i}\in \mathbb{R}^{6N}:\mathbf{\sigma}_{i}=\left(\mathbf{p}_{i},\mathbf{q}_{i}\right), i\in \left[1,N\right]\right\}$ represents a state of the entire $N-$particle system [\ref{Huang 1987}]. A system with perfect accessibility $\Omega_{PA}$ is a pair $\left(\Omega, \Pi\right)$, with $\Pi$ a set whose elements  $\pi$ are called process generators, together with two functions [\ref{Lucia 2001}]:
\begin{equation}
\pi \mapsto \mathcal{S}
\end{equation}
\begin{equation}
\left(\pi^{'},\pi{''}\right) \mapsto \pi^{''}\pi{'}
\end{equation}
where $\mathcal{S}$ is the state transformation induced by $\pi$, whose domain $\mathcal{D}\left(\pi\right)$ and range $\mathcal{R}\left(\pi\right)$ are non-empty subset of $\Omega$.
This assignment of transformation to process generators is required to satisfy the following conditions of accessibility:
\begin{enumerate}
	\item $\Pi\sigma:=\left\{\mathcal{S}\sigma :\pi\in\Pi,\sigma\in \mathcal{D}\left(\pi\right)\right\}=\Omega$ , $\forall \sigma\in\Omega$\emph{:} the set $\Pi\sigma$ is called the set of the states accessible from $\sigma$ and, consequently, it is the entire \emph{state space}, the phase spase $\Omega$;
	\item if $\mathcal{D}\left(\pi ''\right)\cap \mathcal{R}\left(\pi '\right)\neq 0\Rightarrow \mathcal{D}\left(\pi ''\pi '\right)=\mathcal{S}_{\pi '}^{-1}\left(\mathcal{D}\left(\pi ''\right)\right)$ and $\mathcal{S}_{\pi ''\pi '}\sigma =\mathcal{S}_{\pi ''}\mathcal{S}_{\pi '}\sigma$ $\forall\sigma\in \mathcal{D}\left(\pi ''\pi '\right)$
\end{enumerate}
It follows that a process in $\Omega_{PA}$ is a pair $\left(\pi ,\sigma\right)$, with $\sigma$ a state and $\pi$ a process generator such that $\sigma$ is in $\mathcal{D}\left(\pi\right)$. The set of all processes of $\Omega_{PA}$ is denoted by [\ref{Lucia 2001}]:
\begin{equation}
\Pi \diamond\Omega =\left\{\left(\pi ,\sigma\right): \pi \in \Pi ,\sigma\in \mathcal{D}\left(\pi\right)\right\}
\end{equation}
If $\left(\pi ,\sigma\right)$ is a process, then $\sigma$ is called the initial state for $\left(\pi ,\sigma\right)$ and $\mathcal{S}\sigma$ is called the final state for $\left(\pi ,\sigma\right)$. A cycle $\mathcal{C}$ is a path whose endpoints coincide. A thermodynamic system is a system with perfect accessibility $\Omega_{PA}$ with two actions $W\left(\pi ,\sigma\right)\in\mathbb{R}$ and $H\left(\pi ,\sigma\right)\in\mathbb{R}$, called work done and heat gained by the system during the process $\left(\pi ,\sigma\right)$, respectively [\ref{Lucia 2001}]. The set of all these stationary states of a system $\Omega_{PA}$ is called non-equilibrium ensemble [\ref{Lucia 2008}]. Last a thermodynamic path $\gamma$ is an oriented piecewise continuously differentiable curve in $\Omega_{PA}$ [\ref{Lucia 2001}]. The triple $\left(\Omega_{PA},\mathcal{F},\mu_{PA}\right)$ is a \emph{measure space}, the \emph{Kolmogorov probability space} $\Gamma$ [\ref{Lucia 2008}]. A dynamical law $\tau$ is a group of meausure-preserving automorphisms $\mathcal{S}:\Omega_{PA}\rightarrow\Omega_{PA}$ of the probability space $\Gamma$ [\ref{Lucia 2008}]. A dynamical system $\Gamma_{d}=\left(\Omega_{PA},\mathcal{F},\mu_{PA},\tau\right)$ on a measure space $\left(\Omega_{PA},\mathcal{F},\mu_{PA}\right)$ consists of a dynamical law $\tau$ on the probability space $\Gamma$ [\ref{Lucia 2008}].

Following Wang,  we look at a nonequilibrium system moving in the $\Omega$-space between two states which are in two elementary cells of a given partition of the phase space itself. We use the concept of path of classical mechanics: if the motion of the system is regular, or if the phase manyfold has positive or zero riemannian curvature, there will be only a fine bundle of paths which track each other between the initial and the final cells [\ref{Wang 2005}]. These trajectories must be the paths minimizing action according to the principle of least action and have unitary probability; any other path must have zero probability [\ref{Wang 2005},\ref{Biesiada}]. For a system in chaotic motion, or when the riemannian curvature of the phase manyfold is negative, two points indistinguishable in the initial cell can separate from each other exponentially  [\ref{Wang 2005}]. Then between two given phase cells represented there may be many possible paths $\gamma_k$, $k\in [1,\omega]$ with $\omega$ number of all the paths, with different travelling time $t_{\gamma_k}$ of the system and different probability $p_{\gamma_k}$ for the system to take the path $k$. This last quantity is called path probability distribution [\ref{Wang 2005}] and is defined as follows. Consider an ensemble of a large number $L$ of identical systems all moving in the phase space from two cells with $\omega$ possible paths. We observe $L_k$ systems travelling on the path $\gamma_k$, with $k\in [1,\omega ]$. The probability $p_{\gamma_k}$ that the system take the path $\gamma_k$ is defined as usual by $p_{\gamma_k} = L_k /L$. Of course $\sum_{k=1}^{\omega}p_{\gamma_k} =1$. By definition, $p_{\gamma_k}$ is a transition probability from the two states considered [\ref{Wang 2005}]. Wang supposed that the different paths of the non-equilibrium systems moving between the phase cells are uniquely differentiated by their action.

Considering a chaotic open systems, the action has been expressed as follows [\ref{Lucia 1995},\ref{Lucia 2008}]:
\begin{equation}
\mathcal{A}=-T_{ref}\int_{t_{\gamma_k}}\Delta S_{irr}dt
\end{equation}
where $T_{ref}$ is the temperature of the lower reservoir, 
\begin{equation}
\Delta S_{irr}=\frac{Q_{r}}{T_{a}}\left(1- \frac{T_{a}}{T_{r}}\right)+\frac{\Delta H}{T_{a}}-\Delta_{ex} S+\frac{\Delta E_{k}+\Delta E_{g}-W}{T_{a}}
\label{irr}
\end{equation}
is the entropy generation [\ref{Lucia 1995}, \ref{Lucia 1998}-\ref{Lucia 2008}] and $t_{\gamma_k}$ is the travelling time of the system along the path, with $Q_{r}$ is the heat source, $T_{r}$ its temperature, $T_{a}$ the ambient temperature (in general reference temperature, i.e. the least reservoir temperature), $H$ is the enthalpy, $\Delta_{ex}S= S_{in}-S_{out}=\int d_{ex}S$, with $S_{in}$ entropy which enters into the system  [\ref{di Liberto 2007}] and $S_{out}$ entropy which flows out of the system [\ref{di Liberto 2007}], $E_{k}$ the kinetic energy, $E_{g}$ the gravitational energy, $W$ the work done and $W_{lost}$ work lost in irreversibility.

Starting from a dynamical point of view the statistical expression, for the irreversible-entropy variation, has been obtained [\ref{Lucia 2008}]:
\begin{equation}
\Delta S_{irr}=-\frac{k_{B}}{\Dot{m}}\int_{\Omega}\mu_{PA}\left(d\mathbf{\sigma}\right)\sum_{j}\partial_{\sigma_{j}}E_{j}\left(\mathbf{\sigma}\right)
\label{irrentrdef}
\end{equation}
with $\mathbf{E}\left(\mathbf{\sigma}\right)=\mathbf{f}_{int}\left(\mathbf{\sigma}\right)+\mathbf{f}_{nc}\left(\mathbf{\sigma}\right)+\mathbf{f}_{term}\left(\mathbf{\sigma}\right)$, with $\sigma$ states in the perfect accessibility phase space $\Omega_{PA}$, $\mathbf{f}_{int}\left(\mathbf{\sigma}\right)$ internal conservation force, $\mathbf{f}_{nc}\left(\mathbf{\sigma}\right)$ external active non conservative force, $\mathbf{f}_{term}\left(\mathbf{\sigma}\right)$ thermostatic expression and  $\mu _{PA}$ is a statistics such that it describe the asymptotic behaviour of almost all initial data in perfect accessibility phase space $\Omega_{PA}$ such that, except for a volume zero set of initial data $\mathbf{\sigma}$.

For an open thermodynamic system, it has been proved [\ref{Lucia 1995}] the principle of maximum irreversible entropy as a macroscopic and global approach, which generalizes to the open systems the local principle of least entropy production [\ref{3},\ref{4}], giving it a global integration for the open complex systems, as largely discussed in [\ref{Lucia 1995},\ref{LG}]: this integration is the consequence of the difference in a sign, linking to a maximum principle [\ref{Lucia 1995}]. This principle, called also maximum entropy generation principle, states that the condition of stability for the open system is that its irreversible entropy variation $\Delta S_{irr}$ reaches its maximum [\ref{Lucia 2001}-\ref{Lucia 2008}]:
\begin{equation}
\delta\left(\Delta S_{irr}\right)\geq 0
\label{maxent}
\end{equation}

If the paths can be identified only by their actions, then it will be possible to study their probability distributions with the information concept and the method of maximum information of Jaynes in connection with our knowledge about action. This approach will leads us to a probabilistic interpretation of the mechanical principle of Maupertuis and a probability distribution depending on action.

\section{Path information and irreversible entropy}
Since 1962, Jaynes argued that Gibbs' formalism of equilibrium statistical mechanics could be generalised in a statistical inference theory for non-equilibrium systems [\ref{Dewar 2003}]. Moreover,  the fluctuation theory has been established for a variety of non-equilibrium systems: this theorem `relates the probability $p(\sigma_{\Delta t})$ of observing a phase space trajectory with the whole system entropy production rate $\sigma_{\Delta t}$ over time interval $\Delta t$, to that of observing a trajectory with entropy production rate of $-\sigma_{\Delta t}$, specifically $p(\sigma_{\Delta t}) /p(-\sigma_{\Delta t})=\exp(\Delta t\sigma_{\Delta t} /k_B)$' [\ref{Dewar 2003}] which means that the `probability of violation of the Second Law of thermodynamics becomes exponentially small as $\Delta t$ (or the system size) increases' [\ref{Dewar 2003}]. Consequently, Jaynes developed the non-equilibrium statistical mechanics for the stationary steady-state constraint on the basis of maximum entropy; his approach consists of maximising the path Shannon information entropy $S_I=-\sum_\gamma p_\gamma \ln p_\gamma$ with respect to $p_\gamma$ of the path $\gamma$, with the probability subject to the imposed constraints. According to Shannon, `the information entropy is the logarithm of the number of the outcomes $i$ with non-negligible probability $p_i$', while in `non-equilibrium statistical mechanics it is the logarithm of the number of microscopic phase-space paths $\gamma$ having non-negligible probability $p_\gamma$' [\ref{Dewar 2003}]. Jaynes' procedure consists of finding the `most probable \textit{macroscopic path} realised by the greater number of microscopic paths compatible with the imposed constrained' [\ref{Dewar 2003}], in analogy with the Boltzmann microstate counting: `paths rather then states are the central objects of interest in non-equilibrium systems, because of the presence of non-zero macroscopic fluxes whose statistical description requires considering the underlying microscopic behaviour over time' [\ref{Dewar 2003}] which implies that `the macroscopic behaviour is reproducible under given constraints' and it is `characteristic of each of the great number of microscopic paths compatible with those constraints' [\ref{Dewar 2003}]. Then, Jaynes maximised the information entropy with respect to $p_\gamma$, subject to fixed initial configurations of internal energy and mass density $\textbf{d}\big(u,\{\rho_i\},t\big)$, with $\rho$ mass density and $u$ specific energy, within the control volume $V$ at the time $t =0$ and fixed time-averaged configurations of internal energy mass flux density $\textbf{F}\big(\textbf{f}_u,\{\textbf{f}_i\}\big)$, with $\textbf{f}_u(\textbf{x},t)$ internal energy density and $\textbf{f}_i(\textbf{x},t)$ mass flux density. He introduced the Lagrange multipliers $\pmb{\lambda}$ conjugate to $\textbf{d}$ and $\textbf{F}$ and leads to a path distribution and which allows to maximize information entropy $S_{I, max}=-\sum_\gamma p_\gamma \ln p_\gamma$ with respect of $\pmb{\lambda}$. From this last relation the entropy generation can be defined as [\ref{Lucia 1998}-\ref{Lucia 2008}]:
\begin{equation}
\begin{split}
\Delta S_{irr}=-k_B S_{I} = -k_B\sum_k p_{\gamma_{k}} \ln p_{\gamma_{k}}
 \end{split}
 \label{mug}
\end{equation}
This last relation is the value of the integral in the statistical definition of the entropy generation (\ref{irrentrdef}). It can be also interpreted as the missing information necessary for predicting which path a system of the ensemble takes during the transition from a state to another.

\section{Paths and stochastic order}
When $\mu_{PA}\left(\Omega_{PA}\right)$ is finite, then for any integrable function $\varphi :\Omega_{PA}\rightarrow \mathbb{R}$, the time average $\left\langle\varphi\right\rangle_{t}$ on $\gamma$ is defined for all paths $\gamma$ outside of a set $\mathcal{N}_{\varphi}$ of measure $\mu\left(\mathcal{N}_{\varphi}\right)= 0$. Furthermore $\left\langle\varphi\right\rangle_{t}$ is integrable, with $\left\langle\varphi\right\rangle_{t}\circ \gamma = \left\langle\varphi\right\rangle_{t}$ wherever it is defined, and with
\begin{equation}
\int _{\Omega_{PA}}\left\langle\varphi\right\rangle_{t}d\Omega_{PA}=\int _{\Omega_{PA}}\varphi d\Omega_{PA}
\end{equation}
Following the Birkoff's ergodic theorem, a measure preserving transformation $\mathcal{S}:\Omega_{PA}\rightarrow\Omega_{PA}$ is ergodic if and only if, for every integrable function $\varphi:\Omega_{PA}\rightarrow\mathbb{R}$, the time average $\left\langle\varphi\right\rangle_{t}$ on $\gamma$ is equal to the space average $\int_{\Omega_{PA}}\varphi\left(\mathbf{\sigma}\right) \mu_{PA}\left(d\mathbf{\sigma}\right)$ for all points $\mathbf{\sigma}$ outside of some subset $\mathcal{N}_{\varphi}$ of measure $\mu_{PA}\left(\mathcal{N}_{\varphi}\right) = 0$. Consequently, results of the measurements as the infinite time averages of phase functions because they take a long time compared to the microscopic relaxation time and, for metrically  transitive ($=$ ergodic) systems, measurement results are almost always equal to microcanonical averages. Moreover, the initial states lead to different paths in phase space, so the averages depend on the initial state. Let $\Gamma_{d}$ be a dynamical system and $\varepsilon=\left\{\varepsilon_{\mathcal{S}}\right\}$ be fixed and non zero. An open system is in $\varepsilon-$steady state during the time interval $\mathcal{T}$ if and only if for all $\mathcal{S}\in\Omega_{PA}$ there exists $\zeta_{\mathcal{S}}\in\mathbb{R}$ such that for all $t\in\mathcal{T}$ it follows:
\begin{equation}
\left|\left\langle \mathcal{S}\right\rangle_{\mathcal{T}}-\zeta_{\mathcal{S}}\right|\leq \varepsilon_{\mathcal{S}}
\end{equation}
As a consequence of this last definition, the probability may fluctuate within small bounds and, consequently, dynamical evolution towards steady states is allowed. Every probability space $\Gamma$ generates probability algebra with the $\sigma$-complete Boolean algebra $\mathcal{B}=\mathcal{F}/\Delta$, where $\Delta$ is the $\sigma$-ideal of Borel sets of $\mu_{PA}$-measure zero and the restriction of $\mu_{PA}$ to $\mathcal{B}$ is a strictly positive measure $p$. Conversely, every probability algebra $\left(\mathcal{B},p\right)$ can be realized by some Kolmogorov probability space $\Gamma$ [\ref{Lucia 2008}]. The Borel $\sigma$-algebra $\mathcal{F}_{\mathbb{R}}$ of subsets of the set $\mathbb{R}$ of real numbers is the $\sigma$-algebra generated by the open subsets of $\mathbb{R}$. In a Kolmogorov's set-theoretical formulation, a statistical observable is a $\sigma$-homomorphism $\xi :\mathcal{F}_{\mathbb{R}}\rightarrow \mathcal{B}$. Every observable $\mathbf{\xi}$ can be induced by a real-valued Borel function $\pi :\Omega_{PA}\rightarrow \mathbb{R}$ by means of the inverse map [\ref{Primas 1999}]:
\begin{equation}
\mathbf{\xi}\left(\mathcal{R}\right):=\pi^{-1}\left(\mathcal{R}\right):=\left\{\sigma\in\Omega_{PA}:\pi\left(\sigma\right)\in\mathcal{R}\right\}, \mathcal{R}\in\mathcal{F}_{\mathbb{R}}
\end{equation}
A real-valued Borel funcion $\pi$ defined on $\Omega_{PA}$ is said real valued random variable. Moreover, let it be  $\pi :\Omega_{PA}\rightarrow \mathbb{R}$ a real-valued Borel function such that $\mathbf{\sigma}\mapsto\mathcal{S}\left(\mathbf{\sigma}\right)$ is integrable over $\Omega_{PA}$ with respect to $\mu_{PA}$, the expectation value $\pi_{ev}$ of $\mathcal{S}\left(\mathbf{\sigma}\right)$ with respect to $\mu_{PA}$ is:
\begin{equation}
\pi_{ev}:=\int_{\Omega}\mathcal{S}\left(\mathbf{\sigma\mathbf}\right)\mu_{PA}\left(d\mathbf{\sigma}\right)
\end{equation}
A real-valued random variable $\pi :\Omega_{PA}\rightarrow \mathbb{R}$ on $\Gamma$ induces a probability measurement $\mu_{PA}:\mathcal{F}\rightarrow\left[0,1\right]$ on the state space $\left(\mathcal{F}_{\mathbb{R}},\mathbb{R}\right)$ [\ref{Primas 1999}]. Considering the probability as a property of the generating conditions of a sequence, randomness can be related to predictability and retrodictability \emph{(Primas 1999)}. A family $\left\{\mathbb{\xi}\left(t\right):t\in\mathcal{R}\right\}$ is called a stocastic process, which can be represented by a family $\left\{\gamma\left(\mathbf{\sigma}\left(t\right)\right):t\in\mathbb{R}\right\}$ of equivalent classes of random variables $\mathbf{\xi}\left(t\right)$ on $\Gamma$. The point function $\gamma\left(\mathbf{\sigma}\left(t\right)\right)$ is called trajectory of the stocastic process $\mathbf{\xi}\left(t\right)$. The description of physical systems in terms of a trajectory of a stochastic process corresponds to a point dynamics, while its description in terms of equivalent classes of trajectories and an associated probability measure corresponds to an ensamble dynamics [\ref{Primas 1999}]. 

The principle of maximum irreversible entropy (\ref{maxent}), in addition to the statistical expression of the entropy generation (\ref{mug}), induce a stochastic order into the thermodynamic paths in the phase space. Indeed, the principle of maximum irreversible entropy allows us to single out the most probable path in the phase space, followed by the system in its evolution. Let us consider the set $\{\gamma_k, k\in[1,N]\}$ of possible $N$ paths in the phase space. We say that $\gamma_i$ is smaller than $\gamma_j$, for all $i$ and $j$ in $[1,N]$, in the stochastic order, denoted by $\gamma_i <_{ST} \gamma_j, \forall i,j\in[1,N]$, if
\begin{equation}
p_{\gamma_{i}}<p_{\gamma_{j}},  \forall i,j\in[1,N]
\end{equation}

Considering the relation (\ref{mug}) it follows:
\begin{equation}
-\frac{1}{k_B}\frac{\partial \Delta S_{irr}}{\partial p_{\gamma_k}}=1+\ln p_{\gamma_k}
\end{equation}
from which we can obtain:
\begin{equation}
p_{\gamma_k}=\exp\bigg(-\frac{1}{k_B}\frac{\partial \Delta S_{irr}}{\partial p_{\gamma_k}}-1\bigg)
\label{probabi}
\end{equation}
Now, considering that:
\begin{equation}
\gamma_i <_{ST}\gamma_j, \forall i,j \in[1,N] \Rightarrow p_{\gamma_{i}}<p_{\gamma_{j}},  \forall i,j\in[1,N]
\end{equation}
it follows:
\begin{equation}
\exp\bigg(-\frac{1}{k_B}\frac{\partial \Delta S_{irr}}{\partial p_{\gamma_i}}-1\bigg)>\exp\bigg(-\frac{1}{k_B}\frac{\partial \Delta S_{irr}}{\partial p_{\gamma_j}}-1\bigg), \forall i,j \in[1,N] 
\end{equation}
from which it can be obtained:
\begin{equation}
\frac{\partial \Delta S_{irr}}{\partial p_{\gamma_i}} < \frac{\partial \Delta S_{irr}}{\partial p_{\gamma_j}}, \forall i,j \in[1,N] 
\end{equation}
and, consequently, we can state that: let $\{\gamma_k, k\in[1,N]\}$ be the set of possible $N$ paths in the phase space, $\gamma_i <_{ST}\gamma_j$, $\forall i,j \in[1,N]$, if and only if
\begin{equation}
\frac{\partial \Delta S_{irr}}{\partial p_{\gamma_i}}<\frac{\partial \Delta S_{irr}}{\partial p_{\gamma_j}}, \forall i,j\in[1,N]
\label{entrord}
\end{equation}
So we have proved that the quantities $\dfrac{\partial \Delta S_{irr}}{\partial p_{\gamma_i}}$ are ordered and in stochastic order, too. In theory of probability, it is well known that two random variables $X$ and $Y$ are in stochastic order if there exist a random variable $Z$ and functions $\psi_1$ and $\psi_2$ such that $X=\psi_1(Z)$ and $Y=\psi_2(Z)$ with $\psi_1(Z)\leq \psi_2(Z)$ [\ref{stor}]. Then, it follows that the path $\gamma_i \in \{\gamma_k, k\in[1,N]\}$ is smaller than the path $\gamma_j \in \{\gamma_k, k\in[1,N]\}$, for all $i$ and $j$ in $[1,N]$, in the stochastic order, if it is verified the relation (\ref{entrord}).

For an open thermodynamic system, it has been proved the principle of maximum irreversible entropy as a macroscopic and global approach; it states that the condition of stability for the open system is that its irreversible entropy variation $\Delta S_{irr}$ reaches its maximum (\ref{maxent}) [\ref{Lucia 2001}-\ref{Lucia 2008}], but, considering that:
\begin{equation}
\delta\left(\Delta S_{irr}\right) = \sum_{k}\frac{\partial \Delta S_{irr}}{\partial p_{\gamma_{k}}} dp_{\gamma_{k}}=-k_B \sum_{k} (1+\ln p_{\gamma_k}) dp_{\gamma_{k}}
\label{entdiff}
\end{equation}
which is a sum of positive or null terms, so that it is positive or null. Consequently, at the stability the entropy generation is maximum, as the principle of maximum irreversible entropy states. So, the maximum of the entropy generation is a consequence of the stochastic order of the paths in the phase space. Conversely, from a macroscopic point of view, the stochastic order of the paths in the phase space is a consequence of the maximum of the entropy generation at the stability, and the system evolves on the greatest path in the stochastic order.

In stochastic order some closure properties are proved: here we use them in relation to paths. Their proofs are given for random variables in [\ref{stor}]: the paths can be considered as random variables.

Let $\{\gamma_i, i\in[1,N]\}$ and $\{\gamma'_i, i\in[1,N]\}$ be two sets of $N$ paths inside the phase space. If $\gamma_i <_{ST} \gamma'_i, \forall i\in [1,N]$ then:
\begin{equation}
\sum_{i=1}^N \gamma_i <_{ST} \sum_{i=1}^N \gamma'_i
\end{equation}
so the stochastic order of paths is closed under convolutions. Moreover, if the set $\{\gamma_i, i\in[1,N]\}$ is restricted to the stochastic constrain $Y$ such that $[\gamma_j |_Y] <_{ST} [\gamma_k |_Y], \forall j,k \in [1,N]$ then the stocastic order of paths is closed under mixture.

Last, considering the relation (\ref{probabi}) and comparing it with the one obtained by Wang [\ref{Wang 2004}]:
\begin{equation}
p_{\gamma_k} = \frac{1}{Q}\exp(- \eta \mathcal{A}_{\gamma_k})
\end{equation}
we can obtained the expression for the quantities $Q$ and $\eta$; indeed, it follows:
\begin{equation}
Q=\sum_{k=1}^{\omega}\exp(- \eta \mathcal{A}_{\gamma_k})=e
\end{equation}
and
\begin{equation}
\eta = \frac{1}{k_{B}\mathcal{A}_{\gamma_k}}\frac{\partial \Delta S_{irr}}{\partial p_{\gamma_k}}
\end{equation}

From these two last relations it follows that:
\begin{equation}
Q=e=\sum_{k=1}^{\omega}\exp\bigg(- \frac{1}{k_{B}}\frac{\partial \Delta S_{irr}}{\partial p_{\gamma_k}}\bigg)
\end{equation}
which allows us to obtain a physical link from the mathematical quantities and the physical ones.

\section{Conclusions}
The aim of this paper is to introduce the stochastic order in the analysis of the probability distribution attributed to the different paths of chaotic systems moving between two points in the phase space. If there is no chaos, the path information vanishes and there is between two phase cells only a fine bundle of parallel paths which are the paths of least action with unitary probability of occurrence. In the other hand, if the system under consideration is chaotic, there are possible paths with different actions, and consequently there is information. So, following Wang [\ref{Wang 2004}], we conjecture that the path information may be used as a measure of chaos.

Summarizing, a path information is defined for an ensemble of possible paths of chaotic systems moving between two cells in phase space. It is shown that, if we suppose that the different paths are physically identified by their actions, the maximization of the path information leads to a path probability distribution as a function of the action. This result, just obtained by Wang himself, has been related to the maximum entropy generation principle. More, it has been proved that the probability itself is a function of the first derivative of the entropy generation.

Last the stochastic order is introduced in the thermodynamic approach for the open chaotic systems, obtaining that, at the stability, the entropy generation maximum is consequence of the stochastic order of the paths in the phase space, while, conversely, the stochastic order of the paths in the phase space is a consequence of the maximum of the entropy generation at the stability. It has been proved that the system evolves on the greatest path in the stochastic order.

\bibliographystyle{unsrt}
%\bibliography{bibfile}

\end{document}